\documentclass[fleqn,usenatbib]{mnras}

\usepackage{newtxtext,newtxmath}

\usepackage[T1]{fontenc}
\usepackage{ae,aecompl}

\usepackage{graphicx}	
\usepackage{amsmath}	

\title[ULPs as cosmological standard candles?]{New insights into the use of Ultra Long Period Cepheids as cosmological standard candles}

\author[I. Musella et al.]{
Ilaria Musella$^{1}$\thanks{E-mail: ilaria.musella@inaf.it}
Marcella Marconi,$^{1}$
Roberto Molinaro$^{1}$
Giuliana Fiorentino$^{2}$
\newauthor
Vincenzo Ripepi$^{1}$
Giulia De Somma$^{1,3}$
and Maria Ida Moretti$^{1}$
\\
$^{1}$Osservatorio astronomico di Capodimonte, Vicolo Moiariello 16, 80131, Napoli, Italy\\
$^{2}$INAF – Osservatorio Astronomico di Roma, via Frascati 33, I-00078 Monte Porzio Catone, Roma, Italy\\
$^{3}$ Dipartimento di Fisica "E. Pancini", Universit\'a di Napoli "Federico II", Compl. Univ. di Monte S. Angelo, Edificio G, Via Cinthia, 80126 Napoli, Italy
}

\date{Accepted XXX. Received YYY; in original form ZZZ}

\pubyear{2020}

\begin{document}
\label{firstpage}
\pagerange{\pageref{firstpage}--\pageref{lastpage}}
\maketitle

\begin{abstract}
Ultra Long Period Cepheids (ULPs) are pulsating variable stars with a period longer than 80d and have been hypothesized to be the extension of the Classical Cepheids (CCs) at higher masses and luminosities. If confirmed as standard candles, their intrinsic luminosities, $\sim 1$ to $\sim 3$ mag brighter than typical CCs, would allow to reach the Hubble flow and, in turn, to determine the Hubble constant, $H_0$, in one step, avoiding the uncertainties associated with the calibration of primary and secondary indicators. To investigate the accuracy of ULPs as cosmological standard candles, we first collect all the ULPs known in the literature. The resulting sample includes 63 objects with a very large metallicity spread with $12+\log([O/H])$ ranging from 7.2 to 9.2 dex. The analysis of their properties in the $VI$ period-Wesenheit plane and in the color-magnitude diagram (CMD) supports the hypothesis that the ULPs are the extension of CCs at longer periods, higher masses and luminosities, even if, additional accurate and homogeneous data and a devoted theoretical scenario are needed to get firm conclusions. Finally, the three M31 ULPs, 8-0326, 8-1498 and H42, are investigated in more detail. For 8-1498 and H42, we cannot confirm their nature as ULPs, due to the inconsistency between their position in the CMD and the measured periods. For 8-0326, the light curve model fitting technique applied to the available time-series data allows us to constrain its intrinsic stellar parameters, distance and reddening. 
\end{abstract}




\section{Introduction}

The extragalactic distance scale is largely based on the
period-luminosity (PL) relation of CCs, a well known class of Population I ($t<400$ Myr) pulsating variable stars. Their pulsation period typically ranges from 1 to 100 days and their absolute visual magnitudes ($-8<M_V<-2$ mag) are bright enough to make them easily detectable within the Local Group and (thanks to current space observations with the Hubble Space Telescope)  in external galaxies up to $\sim 30$ Mpc \citep{Riess+11}. To reach cosmologically relevant distances, in the so called Hubble flow, secondary distance
indicators, such  as the Tully-Fisher relation and type Ia Supernovae (SNe Ia)
have to be applied. These, in turn, need to be calibrated with primary distance indicators, including not only CCs but also Pop II standard candles, such as the Tip of the Red Giant Branch or RR Lyrae stars. This procedure implies that any systematic error affecting primary and secondary distance indicators reflects onto the Hubble constant evaluation based on the extragalactic distance scale \citep[see e.g.][]{Freedman+01,Altavilla+04,Dicrisci+06,DeSomma+20a}. The derivation of the Hubble constant $H_0$ from the Cepheid-based extragalactic distance scale has recently drawn a renewed attention in the context of a lively debate on the evidence of a tension between Hubble constant evaluations based on the Cosmic Microwave Background and local values based on the cosmic distance scale \citep[see e.g.][and references therein]{Riess+18,Riess+19}. However, notwithstanding theoretical and observational efforts \citep[see e.g.][and references therein]{Freedman+01,Riess+11,Fiorentino+13,Anderson+18,DeSomma+20a} to reduce the uncertainties associated to primary distance indicators, the systematic errors still affecting secondary distance indicators \citep[see e.g.][and references therein]{Altavilla+04,Verde+19} also contributing to the final $H_0$ error budget.

In this context, the possibility to use primary
indicators observable in the Hubble flow would be very important. 
\citet{Bird+09} pointed out the presence of a small number of
variables with long periods ($80\lesssim P \lesssim 210$ d) in nearby forming galaxies (Magellanic
Clouds, NGC 55, NGC 6822, NGC 300), hypothesized to be the
counterparts at higher luminosity and mass of the CCs. Indeed, their
light curves are very similar to the CC ones and
\citet{Bird+09} show (see their figs. 2, 4 and 5) that they appear to
follow the extension at longer periods of the CC period-luminosity (PL) and period-Wesenheit (PW) relations. Thanks to their high luminosity, this class of
variables could represent candidate  primary distance indicators able to reach, in one step, distances of the order 100 Mpc and beyond, in particular with next generation telescopes such as  
the European Extremely
Large Telescope and the James Web Space Telescope. Despite their promising role,  the use of the ULPs as standard candles is
largely debated in the literature due to the small sample of known
ULPs and the particular observational strategy required to follow repeated pulsation cycles at these very long periods.

\citet{Fiorentino+12,Fiorentino+13} analyzed a sample of 37
ULPs with known $V$ and $I$ magnitudes, identified in galaxies with a very
large metallicity range $12+\log(O/H)$ varying from $\sim7.2$ to 9.2 dex. This sample includes the ULPs collected by \citet{Bird+09} in the galaxies LMC, SMC, NGC 55, NGC 300, NGC 6822 and IZw18, the 2 ULPs in M81 by \citet{Gerke+11} and those identified by \citet{Riess+11}  in the framework of the SH0ES project\footnote{“Supernovae and H0 for
the Equation of State” to observe Cepheid variables in galaxies hosting SNe Ia.}, in the galaxies NGC 1309, NGC 3021, NGC 3370, NGC 4536, NGC 5584, NGC 4038 and NGC 4258. \citet{Fiorentino+12,Fiorentino+13} do not confirm the flat $VI$ Wesenheit 
relation obtained by \citet{Bird+09} but find a relation similar to 
that obtained for the LMC Cepheids 
with no significant dependence on metallicity \citep[as expected for the PW in these filters based on theoretical predictions;][]{Fiorentino+07,Bono+08}, but with an unexpected larger spread. 
This spread can be due to many different contributions such as, for example, poor statistics or light curve sampling for some ULPs, the use of non-homogeneous photometric data, but also the adoption of reddening and metallicity values from different sources.

Moreover, the two ULPs identified in the very metal poor
blue compact dwarf galaxy IZw18 have very long periods (about 125 and
130 days, respectively) and are very interesting because, at this very low
metallicity range, evolution and pulsation models do not predict the
existence of such ULPs \citep{Fiorentino+10,Marconi+10}.

\citet{Riess+11} CC samples were enlarged, updated and re-calibrated by \citet{Riess+16} and \citet{Hoffmann+16} finding a sample of 40 ULPs in 14 galaxies. These authors obtained a new calibration for all the observed samples and applied for all the galaxies a consistent procedure to identify variable stars and their properties. Among the 19 ULPs identified by \citet{Riess+11} in the three galaxies NGC 1309, NGC 3021 and NGC 3370 and adopted by \citet{Fiorentino+12}, only 16 were confirmed as ULPs, but with a different period.  

In addition, \citet{Ngeow+15}, using the $R$-band data of the Palomar Transient Factory (PTF), identified a sample of ULP candidates. For these variables, the authors performed a follow-up to obtain $VI$ band time-series, and only two, namely 8-0326 and 8-1498, have been classified as ULPs with period of $74.427 \pm 0.120$ d and $83.181 \pm 0.178$ d, respectively (even if one of these stars has a period shorter than 80 d). Then, they used these variables to derive  M31 distance and test their goodness as standard candles.  

Very recently, \citet{Taneva+20} published  $BVR$ photometry for another candidate M31 ULP, identified using PTF data, namely H42 with a period of 177.32 d.

The resulting sample, including 63 objects, is also characterized by a large metallicity spread and is statistically more significant than those adopted by \citet{Bird+09} and \citet{Fiorentino+12,Fiorentino+13}, allowing us to improve the analysis of the properties of these variables and to get information on their use as standard candles for the cosmic distance ladder.

The ULP sample adopted in this paper is presented in Section \ref{sec:ULPsample}. Their $VI$ Wesenheit relation is discussed in Section \ref{sec:wesulp} and the other ULP properties are analyzed in section \ref{sec:ULPproperties}. In Section \ref{sec:M31ULPs} we discuss the comparison of the ULP sample with theoretical pulsational models, analysing, in particular, the properties of the M31 ULPs, for which time-series data are available \citep{Ngeow+15,Taneva+20},  
and the light curve model fitting technique can be applied \citep[see also][and references therein]{MarconiLC+13,Marconi+17,Ragosta+19}. The Conclusions close the paper.

\section{ULP sample}\label{sec:ULPsample}

In Table \ref{tab:ULPsample}, we list the ULP sample we collected in this work for which we have $VI$ mean magnitudes. It includes the 18 ULPs compiled by \citet{Bird+09} (hereinafter Bird sample), the 2 M81 ULPs by \citet{Gerke+11}, the 2 M31 ULPs confirmed by \citet{Ngeow+15} and the 40 ULPs identified by \citet{Riess+16} and \citet{Hoffmann+16} (hereinafter SH0ES sample).
We do not have new ULPs or new mean magnitude determinations for already known ULPs neither in the Gaia DR2 Cepheid sample reclassified by \citet{Ripepi+19} nor in the recent OGLE Collection of Variable Stars in the MW and the MCs \citep{Sosz+15,Sosz+17,Udalski+18,Sosz+19}.

In Table \ref{tab:ULPsample}, the color excesses are those of the host galaxies based on the Galactic dust reddening maps by \citet{Schlafly+11}.  The references relative to $V$, $V-I$, distance modulus and $12+\log(O/H)$ for the Bird sample and  IZw18 
are reported in \citet{Fiorentino+12} and for M81 in \citet{Gerke+11}. For M31, $V$ and $V-I$ are taken from \citet{Ngeow+15}, the distance modulus from \citet{deGrijs+14}, the metal abundances of the two ULPs using their position \citep{Lee+13} and the metallicity gradient measured in this galaxy by \citet{PHAT_met15}\footnote{\citet{PHAT_met15},
in the framework of the PHAT survey, analysed the metallicity
distribution of 160 Cepheids in the Andromeda galaxy, finding a metallicity $12+\log([O/H])$ varying between 8.82 and 9.12 dex (see their Figure 11) corresponding to a $Z$ ranging between 0.01 and 0.03.} instead of using M31 mean metallicity. For the SH0ES sample, $V$ and $V-I$ are obtained applying photometric transformation by \citet{Sahu+14} to the UVIS-WFC3 F555W and F814W \citet{Hoffmann+16}; the distance moduli and the individual metal abundances (obtained from the metallicity gradient of the host galaxy) are tabulated in \citet{Riess+16} and \citet{Hoffmann+16}. For all the ULPs, we also report the corresponding $Z$ metallicity\footnote{$[O/H]=\log(O/H)-\log(O/H)_{\odot}$ with $\log(O/H)_{\odot}=-3.10$. Assuming that $[O/H]=[Fe/H]$, we obtain $[O/H]=\log Z - \log Z_{\odot}$ and then $Z=10^{[O/H]+\log Z_{\odot}}$, with $Z_{\odot}=0.02$}. 

\begin{table*}
	\centering
	\caption{ULPs with $V$ and $I$ mean magnitudes.}
	\label{tab:ULPsample}
	\begin{tabular}{llccccll} 
		\hline
		Galaxy & Period & $V$ & $V-I$& $\mu_0$ & $E(B-V)^{(1)}$ & $12+\log(O/H)$& $Z$\\
		       &   (d)  &(mag)& (mag)& (mag) & (mag)          & (dex)         &    \\
		\hline
		\multicolumn{8}{c}{Bird Sample} \\
		\hline
LMC     & 109.2  &  12.41  &  1.07  &  18.50 &  0.07   &  8.396  &  $\sim$0.008    \\    
LMC     &  98.6  &  11.92  &  1.11  &  18.50 &  0.07   &  8.396  &  $\sim$0.008    \\    
LMC     & 133.6  &  12.12  &  1.09  &  18.50 &  0.07   &  8.396  &  $\sim$0.008    \\    
SMC     & 210.4  &  12.28  &  0.83  &  18.93 &  0.03   &  7.982  &  $\sim$0.002    \\    
SMC     & 127.5  &  11.92  &  1.03  &  18.93 &  0.03   &  7.982  &  $\sim$0.002    \\    
SMC     &  84.4  &  11.97  &  0.91  &  18.93 &  0.03   &  7.982  &  $\sim$0.002    \\    
NGC55   & 175.9  &  19.25  &  0.84  &  26.43 &  0.01   &  8.053  &  $\sim$0.003    \\    
NGC55   & 152.1  &  19.56  &  0.95  &  26.43 &  0.01   &  8.053  &  $\sim$0.003    \\    
NGC55   & 112.7  &  20.18  &  1.05  &  26.43 &  0.01   &  8.053  &  $\sim$0.003    \\    
NGC55   &  97.7  &  20.54  &  1.25  &  26.43 &  0.01   &  8.053  &  $\sim$0.003    \\    
NGC55   &  85.1  &  20.84  &  1.38  &  26.43 &  0.01   &  8.053  &  $\sim$0.003    \\    
NGC300  & 115.8  &  20.13  &  0.97  &  26.37 &  0.01   &  8.255  &  $\sim$0.004    \\    
NGC300  &  89.1  &  19.71  &  1.02  &  26.37 &  0.01   &  8.255  &  $\sim$0.004    \\    
NGC300  &  83.0  &  19.26  &  0.77  &  26.37 &  0.01   &  8.255  &  $\sim$0.004    \\     
NGC6822 & 123.9  &  17.86  &  1.40  &  23.31 &  0.21   &  8.114  &  $\sim$0.003    \\    
IZw18   & 130.3  &  23.96  &  0.96  &  31.30 &  0.03   &  7.211  &  $\sim$0.0004   \\    
IZw18   & 125.0  &  23.65  &  0.97  &  31.30 &  0.03   &  7.211  &  $\sim$0.0004   \\  
		\hline 
		\multicolumn{8}{c}{M81 ULPs}\\
		\hline 
M81     & 96.8   &  21.52  &  1.40  &  27.69 &  0.07   &  8.77   &  $\sim$0.013    \\    
M81     & 98.981 &  21.69  &  1.42  &  27.69 &  0.07   &  8.77   &  $\sim$0.013    \\    
		\hline 
		\multicolumn{8}{c}{M31 ULPs}\\
		\hline 
M31     & 74.427 &  18.684 &  1.428 &  24.46 &  0.05   &  9.03   &  $\sim$0.02     \\    
M31     & 83.181 &  18.856 &  1.073 &  24.46 &  0.05   &  9.03   &  $\sim$0.02     \\    
\hline 
	\end{tabular}
\end{table*}		
\begin{table*}
	\centering
	\contcaption{ULPs with $V$ and $I$ mean magnitudes.}
	\begin{tabular}{llccccll} 
		\hline
		Galaxy & Period & $V$ & $V-I$& $\mu_0$ & $E(B-V)^{(1)}$ & $12+\log(O/H)$& $Z$\\
		       &   (d)  &(mag)& (mag)& (mag) & (mag)          & (dex)         &    \\
		 \hline      
\multicolumn{8}{c}{SH0ES sample}\\
		\hline 
M101     & 81.521 & 22.70 & 1.04 & 29.14 & 0.008 & 9.15  &  0.028 \\ 
NGC1015  & 87.327 & 25.90 & 1.04 & 32.50 & 0.029 & 8.704 &  0.010 \\ 
NGC1015  & 97.489 & 26.09 & 1.15 & 32.50 & 0.029 & 9.033 &  0.022 \\ 
NGC1309  & 80.886 & 25.87 & 1.10 & 32.52 & 0.035 & 9.115 &  0.026 \\ 
NGC1309  & 84.543 & 26.89 & 1.03 & 32.52 & 0.035 & 8.885 &  0.015 \\ 
NGC1309  & 84.888 & 26.00 & 1.23 & 32.52 & 0.035 & 9.007 &  0.020 \\ 
NGC1309  & 90.592 & 26.54 & 1.27 & 32.52 & 0.035 & 8.781 &  0.012 \\ 
NGC1309  & 90.713 & 26.37 & 1.23 & 32.52 & 0.035 & 8.838 &  0.014 \\ 
NGC1309  & 90.911 & 26.51 & 1.02 & 32.52 & 0.035 & 9.061 &  0.023 \\ 
NGC1448  & 93.353 & 25.08 & 1.15 & 31.31 & 0.012 & 8.852 &  0.014 \\ 
NGC1448  & 97.203 & 25.32 & 1.37 & 31.31 & 0.012 & 8.849 &  0.014 \\ 
NGC2442  & 81.839 & 27.90 & 1.83 & 31.51 & 0.179 & 9.076 &  0.024 \\ 
NGC2442  & 91.57  & 26.64 & 1.56 & 31.51 & 0.179 & 8.878 &  0.015 \\ 
NGC3370  & 84.917 & 26.03 & 1.05 & 32.07 & 0.028 & 9.029 &  0.021 \\ 
NGC3370  & 88.165 & 25.51 & 0.93 & 32.07 & 0.028 & 8.756 &  0.011 \\ 
NGC3370  & 96.096 & 25.84 & 1.16 & 32.07 & 0.028 & 8.798 &  0.013 \\ 
NGC3972  & 85.622 & 25.06 & 1.00 & 31.59 & 0.013 & 8.878 &  0.015 \\ 
NGC3982  & 83.302 & 24.94 & 0.80 & 31.74 & 0.012 & 9.074 &  0.024 \\ 
NGC4038  & 80.257 & 24.03 & 1.08 & 31.29 & 0.041 & 9.046 &  0.022 \\ 
NGC4038  & 80.274 & 25.99 & 1.24 & 31.29 & 0.041 & 9.065 &  0.023 \\ 
NGC4038  & 83.753 & 24.42 & 0.88 & 31.29 & 0.041 & 9.105 &  0.025 \\ 
NGC4038  & 93.069 & 25.42 & 1.21 & 31.29 & 0.041 & 9.055 &  0.023 \\ 
NGC4038  & 93.35  & 24.53 & 0.90 & 31.29 & 0.041 & 9.01  &  0.020 \\ 
NGC4038  & 93.573 & 25.68 & 1.45 & 31.29 & 0.041 & 8.937 &  0.017 \\ 
NGC4038  & 94.396 & 25.38 & 0.99 & 31.29 & 0.041 & 9.026 &  0.021 \\ 
NGC4038  & 95.644 & 24.06 & 0.82 & 31.29 & 0.041 & 9.071 &  0.024 \\ 
NGC4038  & 97.11  & 24.35 & 0.70 & 31.29 & 0.041 & 9.094 &  0.025 \\ 
NGC4258  & 83.258 & 23.20 & 1.07 & 29.39 & 0.014 & 8.743 &  0.011 \\ 
NGC4258  & 84.618 & 23.60 & 1.42 & 29.39 & 0.014 & 8.77  &  0.012 \\ 
NGC4536  & 93.621 & 24.15 & 0.97 & 30.91 & 0.016 & 8.905 &  0.016 \\ 
NGC4536  & 98.775 & 24.29 & 1.24 & 30.91 & 0.016 & 8.887 &  0.015 \\ 
NGC4639  & 81.011 & 26.35 & 1.39 & 31.53 & 0.023 & 9.055 &  0.023 \\ 
NGC5584  & 81.2   & 25.73 & 1.15 & 31.79 & 0.035 & 8.95  &  0.018 \\ 
NGC5584  & 81.356 & 25.58 & 1.10 & 31.79 & 0.035 & 8.743 &  0.011 \\ 
NGC5584  & 85.106 & 25.18 & 0.98 & 31.79 & 0.035 & 8.836 &  0.014 \\ 
NGC5584  & 85.709 & 25.70 & 1.03 & 31.79 & 0.035 & 8.891 &  0.016 \\ 
NGC5584  & 88.513 & 25.95 & 1.19 & 31.79 & 0.035 & 8.804 &  0.013 \\ 
NGC5584  & 97.752 & 26.18 & 1.42 & 31.79 & 0.035 & 8.811 &  0.013 \\ 
NGC7250  & 83.098 & 25.96 & 1.29 & 31.50 & 0.136 & 8.605 &  0.008 \\ 
UGC9391  & 82.992 & 27.20 & 1.26 & 32.92 & 0.009 & 8.946 &  0.018 \\
\hline 
	\end{tabular}
\end{table*}

For the M31 ULP H42, we do not have any I band measurement, but we have the mean magnitudes in the $B$ and $V$ bands obtained by \citet{Taneva+20}: $V= 18.16$ mag and $(B-V) = 1.32$ mag with a period of $177.32$ d. As for the other M31 ULPs, we adopted  $\mu =24.46$ mag and $E(B-V)= 0.05$ mag for the distance modulus and reddening and determined the metallicity from its position and the metallicity gradient by \citet{PHAT_met15}, obtaining $Z=0.01$.

\section{Period-Wesenheit relations for ULPs}\label{sec:wesulp}

\begin{figure}
	\includegraphics[width=\columnwidth]{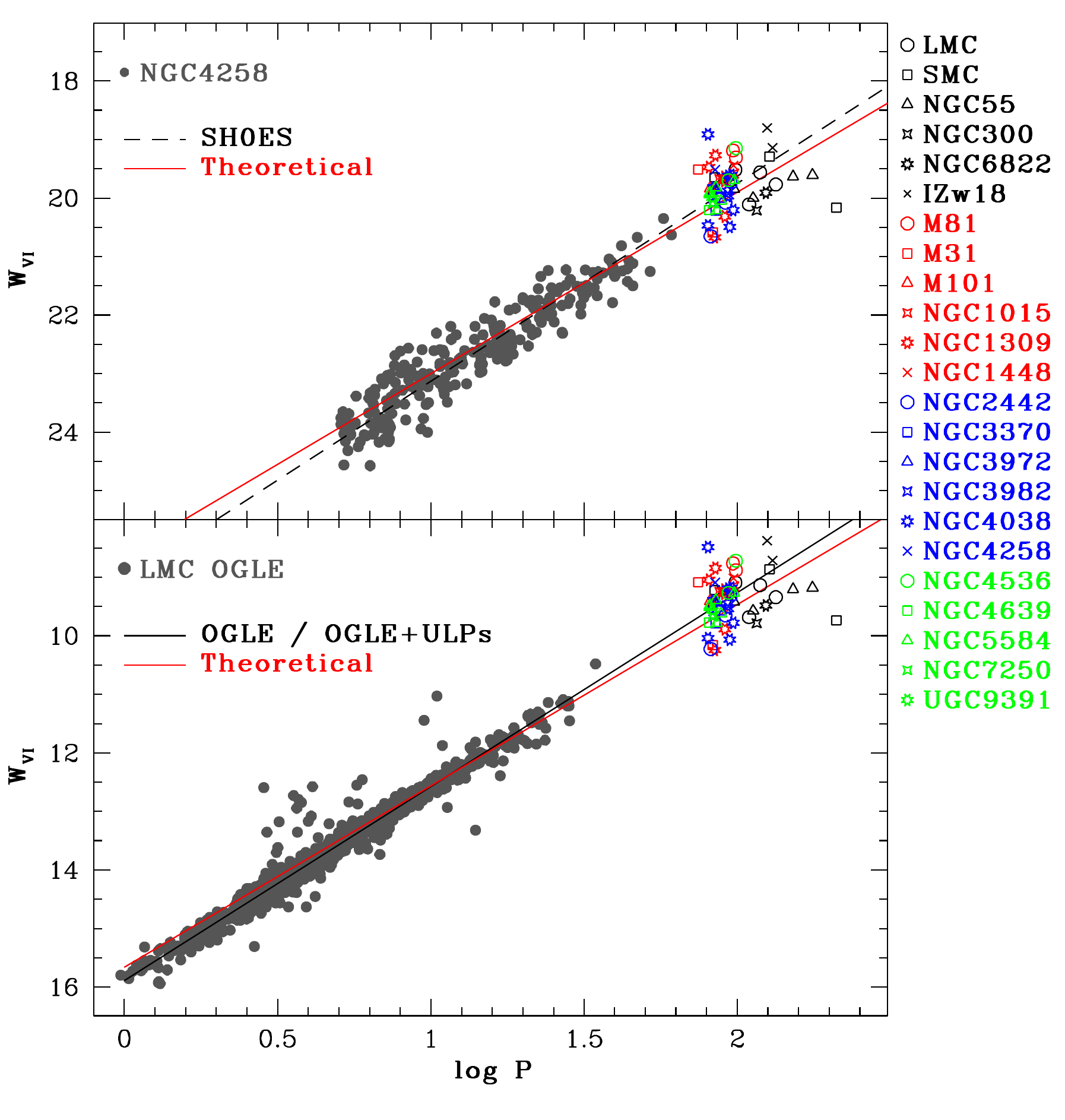}
    \caption{$W_{VI}$ for the ULPs in Table \ref{tab:ULPsample} compared with LMC OGLE  \citep[grey dots in the bottom panel][]{Sosz+15} and NGC 4258 \citep[grey dots in the upper panel][]{Riess+16,Hoffmann+16} CC sample, respectively. The black line in the bottom panel represents the LMC VI Wesenheit obtained by \citet{Sosz+15}, whereas the dashed black line in the upper panel is the VI Wesenheit relation by \citet{Riess+16} in the framework of the SH0ES project. The red line in both panels represents the theoretical metal dependent $W_{VI}^T$ by \citet{Fiorentino+07} adopting $Z=0.01$ (see Sect. \ref{sec:M31ULPs} for details). The symbols adopted for the ULPs are labelled in figure.
    }
    \label{fig:wes_lmc_4258}
\end{figure}

In Fig. \ref{fig:wes_lmc_4258} the ULPs reported in Table \ref{tab:ULPsample} are compared with LMC OGLE  \citep[bottom panel,][]{Sosz+15} and NGC 4258 \citep[upper panel,][]{Riess+16,Hoffmann+16} CCs, respectively, in the P$W_{VI}$ plane, with $W_{VI}=I-1.55(V-I)$. The black line in the bottom panel and the dashed black line in the upper panel represent the $W_{VI}$ by \citet{Sosz+15} (with a slope of $-3.314 \pm 0.008$) and \citet{Riess+16} (with a slope of $-3.38 \pm 0.02$ mag for $P>10$ d obtained with a global fit), respectively.  

The dispersion of the ULPs, in this plot, is much larger than that of the LMC OGLE CCs, but very similar to that of the NGC~4258 Cepheid sample. There are many possible causes for this large spread.  As we know, ULPs, as well as CCs, are observed in very dense environments so that they are subject to  high (and possibly spatially varying) reddening and blending effects. 
A possible variation of the reddening law can influence the spread of the Wesenheit due to a not complete correction of mean (or differential) reddening. However, the very narrow OGLE Cepheids Period-Wesenheit relation seems to support the reliability of the adopted color term. On the other hand, blending can have an important role in the dispersion of this relation. This effect has been analyzed by several authors. In particular, \citet[][and references therein]{Anderson+18} pointed out that the blending effect is the major error source in the distance scale calibration and increases with the distance of the observed galaxies. In our analysis, another possible contribution to the spread can be due to the adopted individual values of reddening and distance. However, a similar spread is observed in much farther galaxies such as NGC 4258, where these effects are not expected to contribute.

\begin{figure}
	\includegraphics[width=\columnwidth]{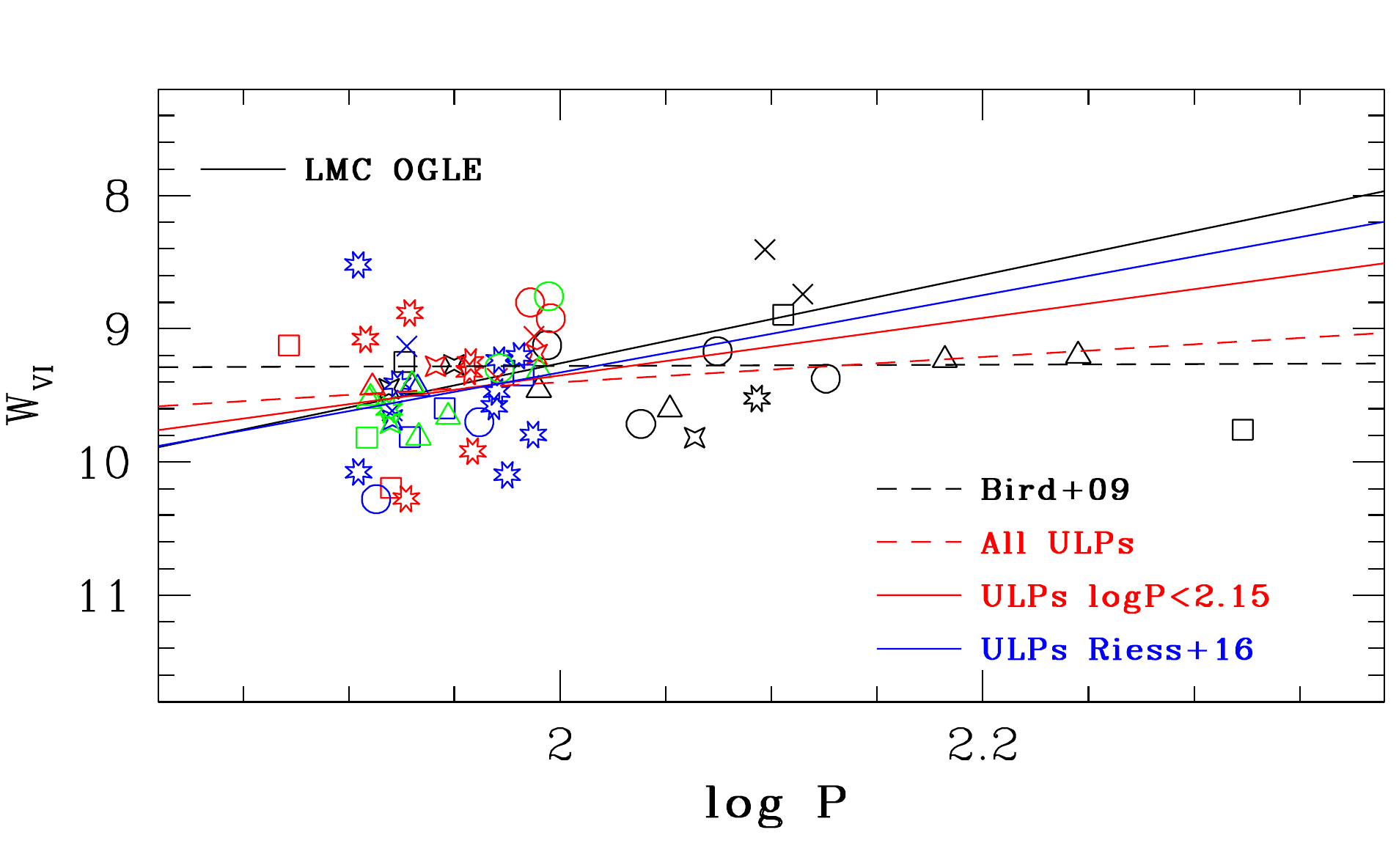}
    \caption{$VI$ Wesenheit function only for the ULPs in Table \ref{tab:ULPsample} placed at the distance of the LMC. The black line is the LMC $W_{VI}$ by OGLE \citep{Sosz+15}, the dashed black line is the  $W_{VI}$ by \citet{Bird+09}, the dashed red line, the red line and the blue line are the $W_{VI}$ obtained in this work by using all the ULPs, those with $logP<2.15$ and only those compiled by \citet{Riess+16}, respectively. The symbols for the ULPs are the same adopted in Fig. \ref{fig:wes_lmc_4258}.
    }
    \label{fig:wesulp}
\end{figure}

\begin{figure}
	\includegraphics[width=\columnwidth]{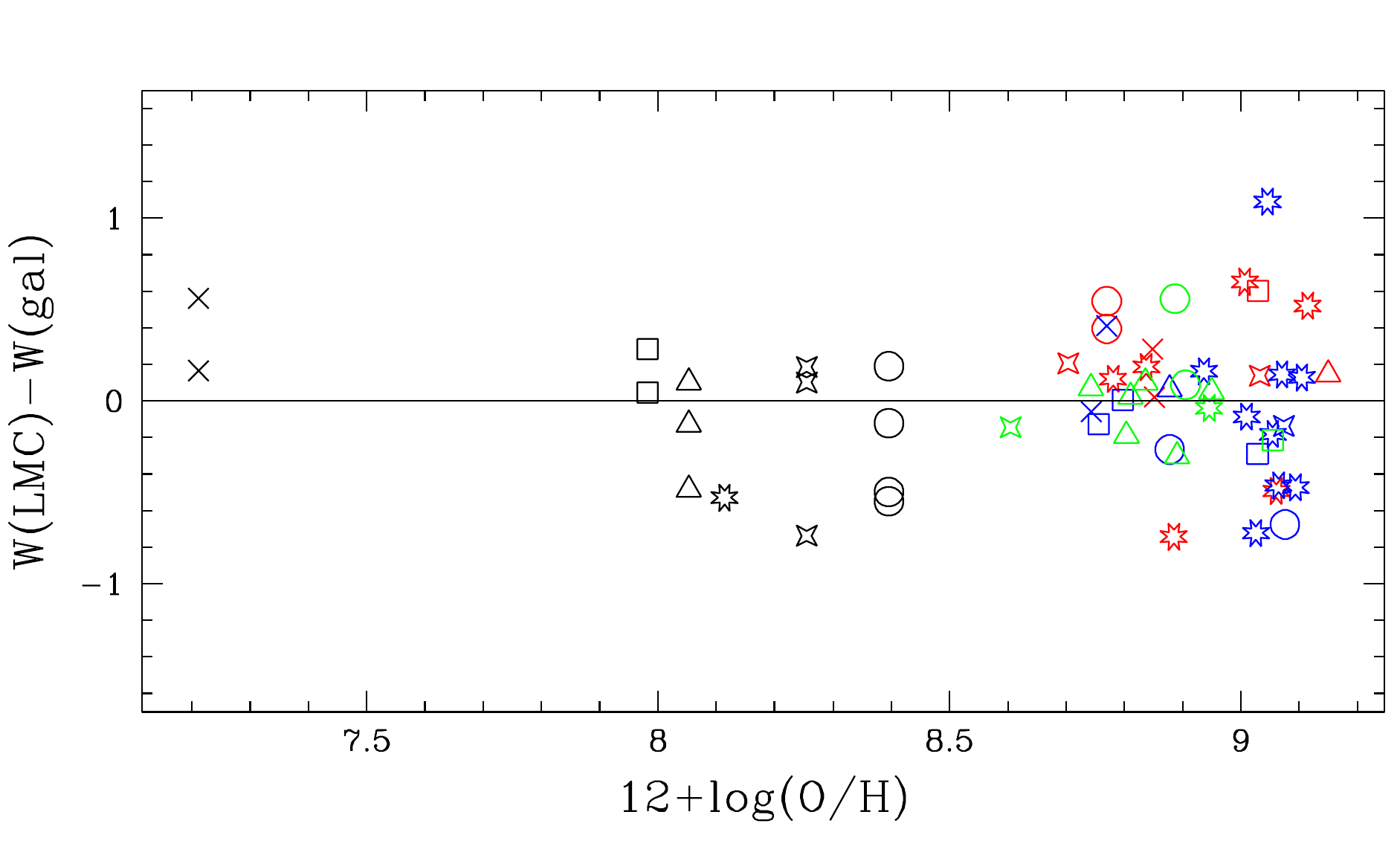}
    \caption{Differences between the Wesenheit relation as defined for the LMC CCs, $W(LMC)$, and that in the host galaxy $W(gal)$ for all the ULPs in Table \ref{tab:ULPsample} versus their metallicity. The symbols for the ULPs are the same adopted in Fig. \ref{fig:wes_lmc_4258}.
    }
    \label{fig:wes_OH}
\end{figure}

In Fig. \ref{fig:wesulp} we compare the distribution in the P$W_{VI}$ plane of the ULPs in Table \ref{tab:ULPsample} with the Period-Wesenheit relations obtained by \citet{Sosz+15} for the LMC OGLE sample (black line), by \citet{Bird+09} for the ULPs (dashed black line), whereas the red dashed line is the relation obtained in this work adopting all the compiled ULPs ($W_{VI} = -0.93 \log P -7.28$ with $\sigma=0.38$). The adoption of a sample collected by different authors can introduce systematic errors in our analysis. For this reason, we carried out a fit for the Wesenheit relation, based only on the SH0ES ULP sample, obtaining $W_{VI} = -2.89 \log P -3.42$ with  an intrinsic dispersion $\sigma=0.36$ (blue line in Fig. \ref{fig:wesulp}). This result is significantly different from the relation obtained adopting the full sample, but we have to note that the range $\log P>2.15$ is very poorly sampled (due to the difficulty to identify and characterize very long period variables), probably creating a false trend for the Wesenheit relation. For this reason, we perform an additional fit excluding these longest periods stars, obtaining $W_{VI} = -2.15 \log P -4.89$ with $\sigma=0.38$ mag (the red solid line in figure). This relation is much steeper than the almost flat slope by \citet{Bird+09}: $W_{VI}=-0.05 \log P - 9.12$ with $\sigma=0.36$ mag and in better agreement with the result based on the SH0ES sample. 
It is worth to note that, in the period range covered by the ULPs, this relation is in good agreement (with a difference in the inferred $W$ magnitude smaller than $\sim 0.05$ mag) with the result by \citet{Sosz+15} for LMC OGLE CCs. On this basis, we perform a global fit including the LMC OGLE CCs and our ULPs, obtaining a slope of $-3.30 \pm 0.01$  and a $\sigma = 0.14$ mag in perfect agreement with that by \citet{Sosz+15}, $-3.314 \pm 0.008$, both using all the ULPs with $\log P \leq 2.15$ and involving only the SH0ES ULPs. The very small error of 0.01 mag, obtained in the global fit, is due to the very large OGLE sample (2455 CCs) that dominates the dispersion term. In any case, this result represents an important hint to consider the ULPs as the counterparts of the CCs at higher mass and luminosity and seems to confirm the higher robustness of the fit obtained using only the SH0ES sample. On the other hand, we have to underline that also this last fit shows a large sigma due to the dispersion of the ULPs around the Wesenheit relation, notwithstanding the accurate and homogeneous photometry. To investigate if a possible origin of this dispersion is the metallicity spread, in Fig. \ref{fig:wes_OH} we show the differences between the Wesenheit relation as defined for the LMC CCs, $W(LMC)$, and that in the host galaxy $W(gal)$ versus the individual metal abundances for all the ULPs with $\log P \leq 2.15$. No significant trend is noted in this plot. The observed spread can be intrinsic and/or due to different effects, such as photometric errors, crowding, blending, but no firm conclusion can be drawn on the basis of current data-sets.

\section {ULP Properties}\label{sec:ULPproperties}

\begin{figure}
	\includegraphics[width=\columnwidth]{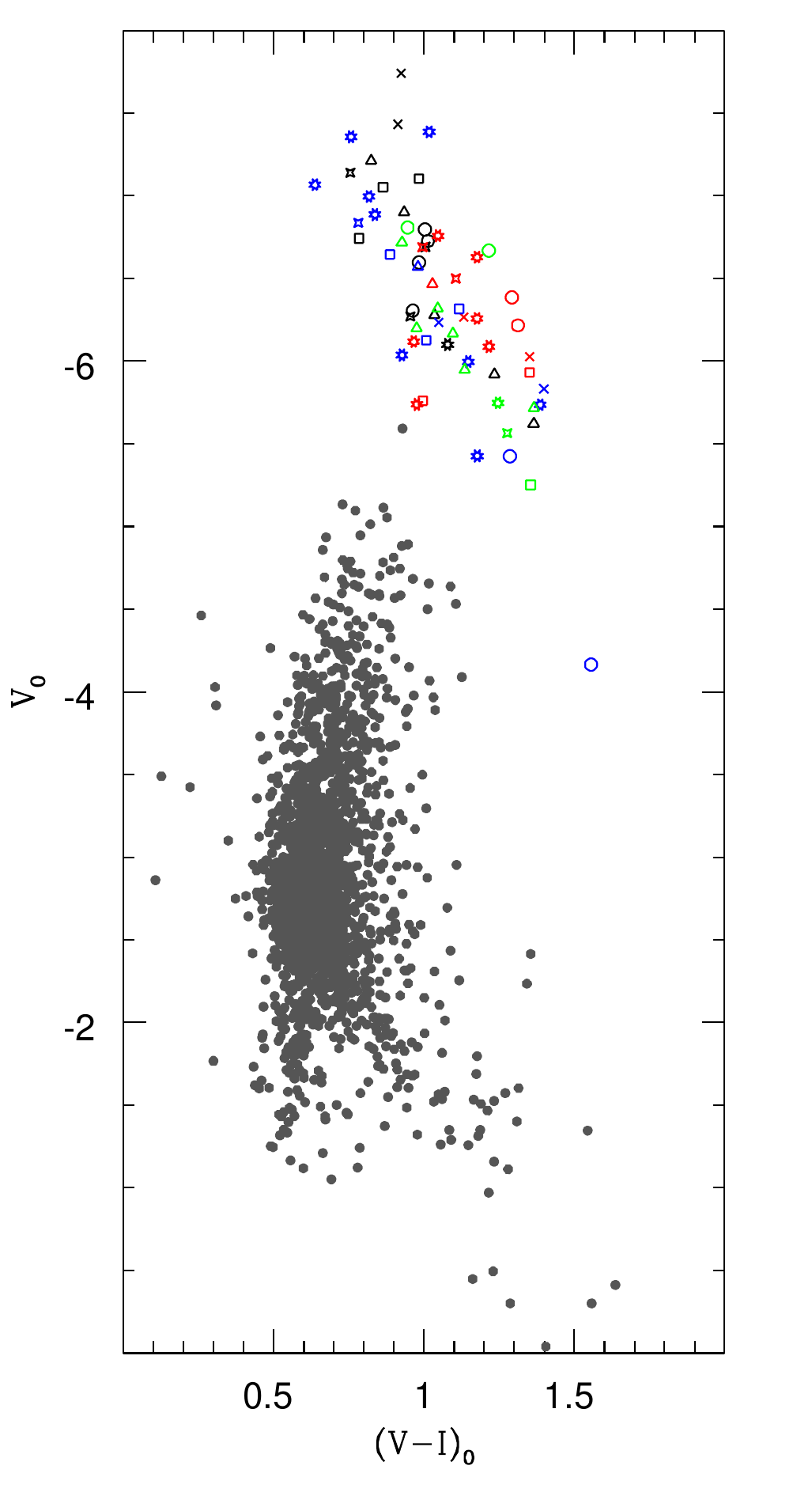}
    \caption{CMD $V_0$ versus $(V-I)_0$ for LMC OGLE Cepheids (grey dots) and ULPs in Table \ref{tab:ULPsample}. The symbols for the ULPs are the same adopted in Fig. \ref{fig:wes_lmc_4258}.
    }
    \label{fig:cmd_cep}
\end{figure}

\begin{figure}
	\includegraphics[width=\columnwidth]{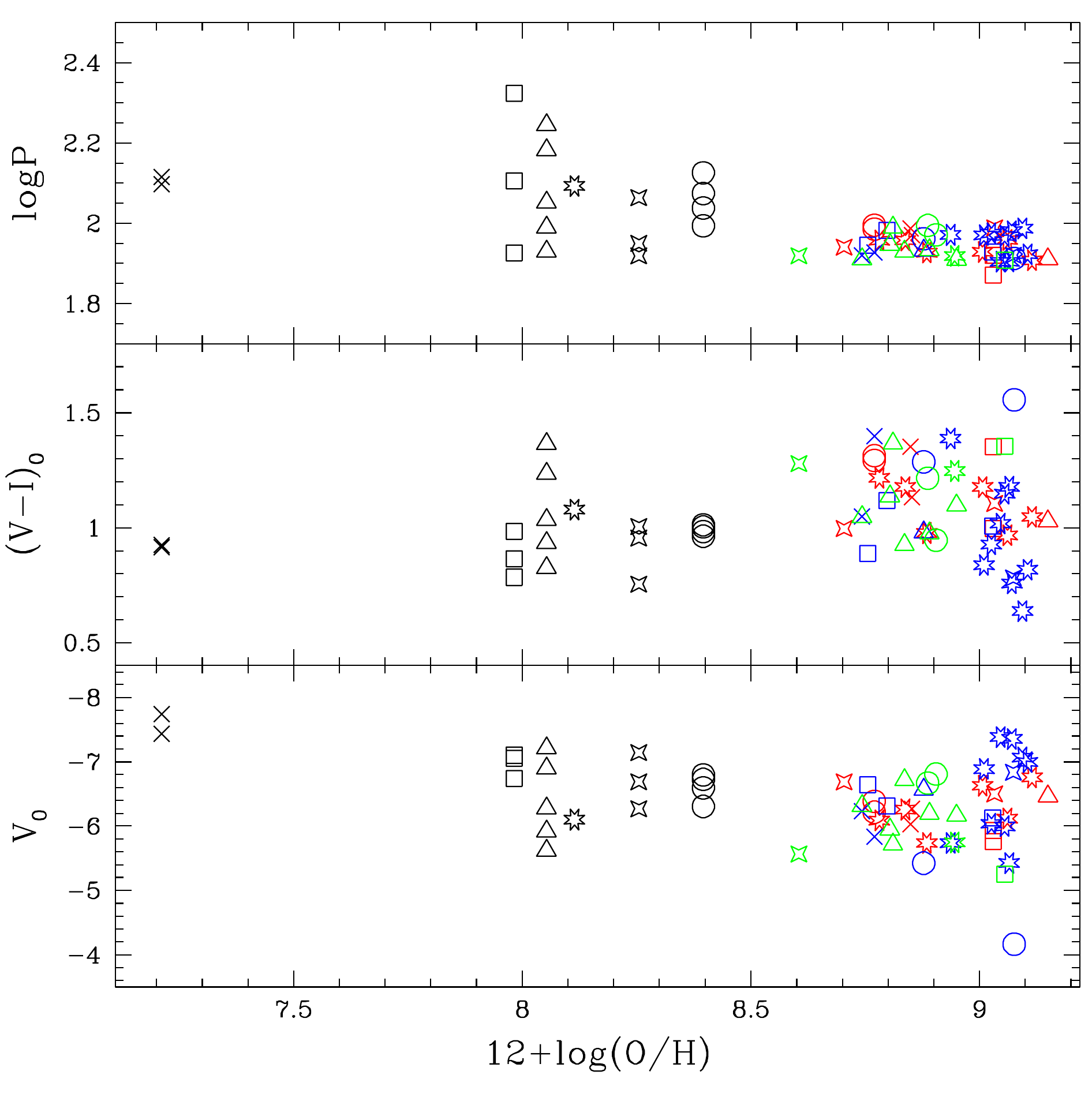}
    \caption{We plot the period (upper panel), and the absolute color (middle panel) and magnitude (bottom panel) of ULPs in Table \ref{tab:ULPsample} versus their metallicity. The symbols for the ULPs are the same adopted in Fig. \ref{fig:wes_lmc_4258}.
    }
    \label{fig:OH}
\end{figure}

Fig. \ref{fig:cmd_cep} shows the position of the ULP sample in Table \ref{tab:ULPsample} (the symbols are the same adopted in Fig. \ref{fig:wes_lmc_4258}) in the color-magnitude diagram (CMD) $V_0$ versus $(V-I)_0$, compared with the LMC OGLE CCs \citep[grey dots,][]{Sosz+15}. The ULPs seem to locate in a region that corresponds to the extrapolation of CC Instability Strip towards higher masses and luminosities, thus confirming the result already found in the PW plane, even if a number of objects appear to be more luminous and bluer than expected. To investigate the causes of this behaviour, we analyze in Fig. \ref{fig:OH} the dependence of ULP periods, colors and absolute visual magnitudes on the metal abundance. We notice that more metal-poor ULPs appear to have longer periods and to be slightly brighter and bluer than the other pulsators. This occurrence could partially justify the position in the CMD of the ULPs belonging to SMC, NGC55, NGC300 and IZw18. On the other hand, the SH0ES metal-rich ULPs do not appear systematically redder and fainter, being distributed on a wide range of colors and magnitudes. This holds in particular for the NGC 4038 ones, for which the stellar metallicity has been recently confirmed to be solar by \citet{Lardo+15}. At this stage, we cannot conclude if this behaviour is an intrinsic property of the ULPs or due to a not sufficient photometric accuracy, crowding and/or blending effects. Probably, a more statistically significant sample of photometrically homogeneous and accurate data, covering larger period and metallicity ranges,  is needed to clarify the observed trend.

\begin{figure}
	\includegraphics[width=\columnwidth]{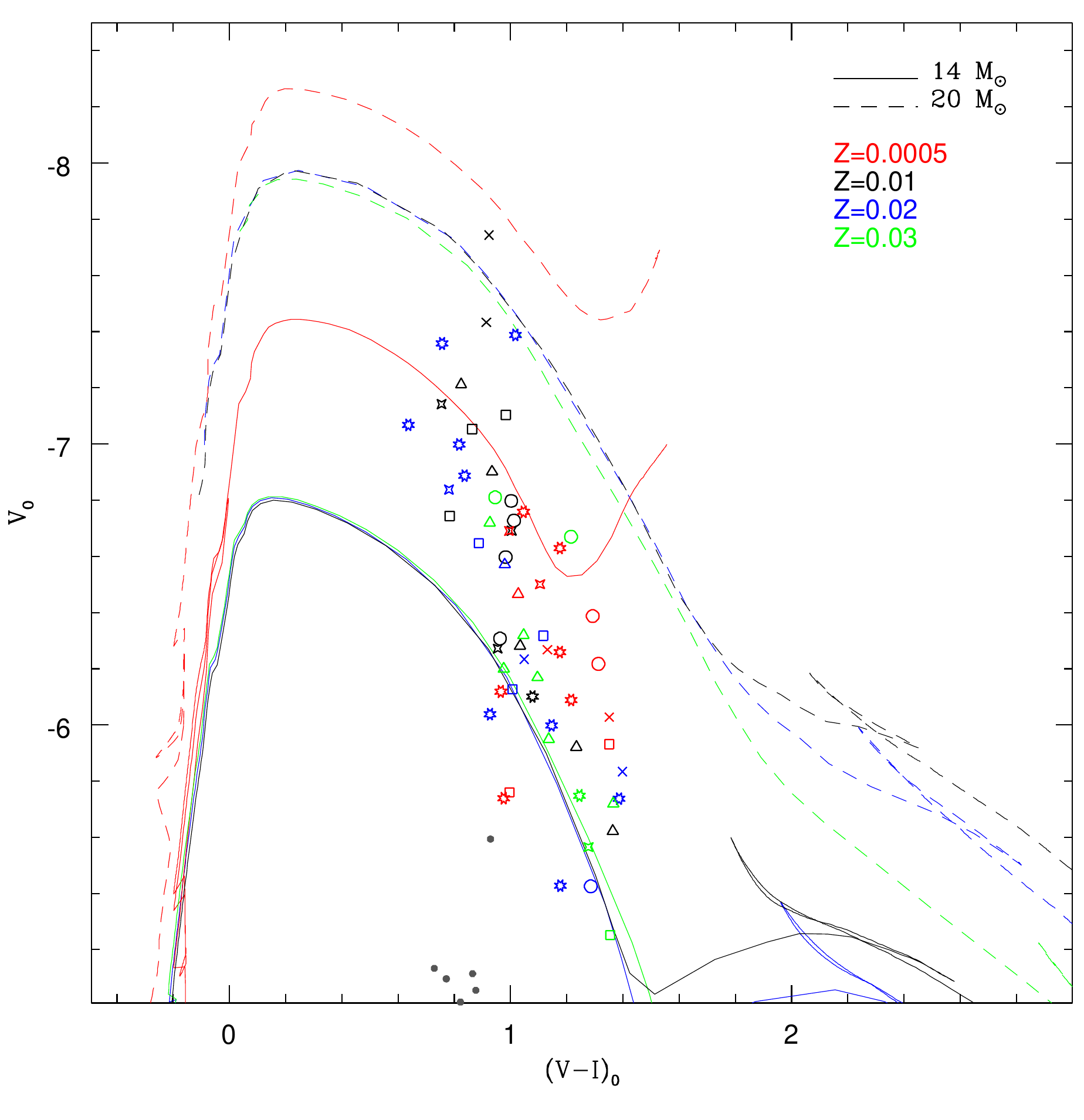}
    \caption{CMD $V_0$ versus $(V-I)_0$ for our ULP sample (the symbols for the ULPs are the same adopted in Fig. \ref{fig:wes_lmc_4258}). The stellar tracks for 14 (solid line) and 20 (dashed line) $M_{\odot}$ and for different metallicities ranging between $Z=0.0005$ to 0.03 (see labels in figure).
    }
    \label{fig:cmd_tracks}
\end{figure}

In Fig. \ref{fig:cmd_tracks} we show the $V_0-(V-I)_0$ CMD only for the ULPs in Table \ref{tab:ULPsample} and the evolutionary tracks by \citet{Bressan+12}\footnote{transformed to the Johnson bands by using the \citet{Chen+19} web tool http://stev.oapd.inaf.it/YBC/index.html}, for 14 (solid lines) and 20 $M_{\odot}$ (dashed lines), including the mass range covered by the ULP luminosities \citep{Bird+09,Fiorentino+12},  and for metallicities ranging from $Z=0.0005$ to $Z=0.03$ (see labels in Fig.\ref{fig:cmd_tracks}), representative of the observed range covered by our sample.
Due to the higher masses and in turn, shorter evolutionary times expected for the ULPs\footnote{Taking into account the very recent theoretical Instability Strip computed by \citet{DeSomma+20a}, we have that the crossing time for a 20 $M_{\odot}$ are about 1.2 Myr and for a 14 $M_{\odot}$, about 2 Myr in comparison with that of $10^5$ and $10^4$ years for a 6 $M_{\odot}$ and a 11 $M_{\odot}$ \citep{DeSomma+20b}, respectively.}, the probability to observe this class of pulsators in an external galaxy is lower than for CCs. Indeed,
the evolutionary tracks for these masses do not show the blue loop crossing the Instability Strip as in the lower mass range.

\section{Comparison with pulsational theoretical models} \label{sec:M31ULPs}

In this section, we compare the observed pulsation properties of the investigated ULPs with the predictions of nonlinear convective pulsation models \citep[see e.g.][and references therein]{Caputo+00,Fiorentino+07,Marconi+05,Marconi+10}. In both panels of Fig. \ref{fig:wes_lmc_4258}, we also plot 
the theoretical metal-dependent Wesenheit relation obtained by \citet{Fiorentino+07}, $W_{VI}^T=-2.67-3.1\log P + 0.08\log(Z)$ with a $\sigma=0.11$ mag, 
shifted for the distance modulus of LMC and NGC4258, respectively. The metallicity adopted to plot this theoretical relation is $Z=0.01$, close to the LMC metallicity, but, as shown in \citet{Caputo+00} and \citet{Fiorentino+07}, the theoretical Wesenheit function built using $V$ and $I$ bands has a negligible dependence on the chemical composition varying by 0.04 mag in the range from $Z=0.01$ to $Z=0.03$. This relation has been obtained in the framework of a theoretical scenario based on non linear, non-local time dependent convective pulsation models relying on  physical and numerical assumptions discussed in our previous papers \citep[see e.g.][and references therein]{Bono+99,Marconi+05} and assuming a large range of
masses (from 3 to 13 $M_\odot$) and chemical compositions ($0.0004 <
Z<0.04$, $0.25<Y<0.33$, \citealp{Fiorentino+02,Marconi+05,Marconi+10}, and references therein).
These models allow us to predict all the relevant pulsation observables, including the period, amplitude and morphology of light and radial velocity curves as a function of the input parameters \citep[see e.g.][and references therein]{Natale+08,Marconi+17}. 

As the theoretical and observational PW relations show a good agreement, within the respective $\sigma$, we can confirm the conclusions reached above, adopting the relation by \citet{Sosz+15}. On the other hand, we deduce that to better understand the ULP behaviour both in the Wesenheit plane and in the CMD, we need to extend our theoretical scenario to the larger masses typical of the ULPs. 

An alternative route to constrain the individual distances, allowing us to simultaneously constrain the intrinsic stellar properties, of pulsating stars is the model fitting of observed light curves \citep[see e.g.][and references therein]{Natale+08,MarconiLC+13,Ragosta+19}. 

Among the ULPs of our sample, for the three M31 ones, we also have time-series data by \citet{Ngeow+15} and \citet{Taneva+20} and the investigation of their light variations could offer a unique opportunity to obtain fundamental information both on the reliability of the adopted pulsation models and on the use of the ULPs as standard candles.

The distance moduli, obtained by applying the theoretical $VI$ Wesenheit relation to the two M31 ULPs found by \citet{Ngeow+15} (the two open squares in the CMD), are $24.07 \pm 0.11$ mag for 8-0326 and $25.29 \pm 0.11$ mag for 8-1498 (the errors are determined by the intrinsic dispersion of the theoretical relation). The obtained average distance modulus is  $24.70 \pm 0.16$ mag, with a large error due to the significant difference between the two individual values, but still consistent with some of the most recent reliable estimates of M31 distance in the literature. In particular, \citet{deGrijs+14} found $24.46\pm0.10$ mag performing an accurate weighted mean of different results obtained by stellar distance indicators, such as Cepheids, RR Lyrae and the tip of the Red Giant Branch, whereas \citet{PHAT15}, using a Cepheid PL relation, found a distance modulus of $24.32\pm 0.09$ mag in the optical bands and $24.51\pm 0.08$ mag in the near-infrared ones. 

Based on the results in section \ref{sec:wesulp}, we derive the distance moduli also applying the ULP Wesenheit relation obtained using only the SH0ES sample. The resulting distance moduli are $24.38 \pm 0.36$ mag and $25.51\pm 0.36$ mag (the errors are determined by the $\sigma$ of the relation) for 8-0326 and 8-1498, respectively. In this case, the mean distance is $24.94 \pm 0.51$ mag, consistent within the errors with the other quoted M31 distance evaluations, but with a much larger error.

Both adopting the theoretical relation and the empirical one based on SH0ES ULPs, we find a large deviation of the distance modulus of 8-1498. The peculiar properties of this star are discussed in the following.

\subsection{Model fitting of the M31 ULPs} \label{sec:modelfitting}

\begin{figure}
	\includegraphics[width=\columnwidth]{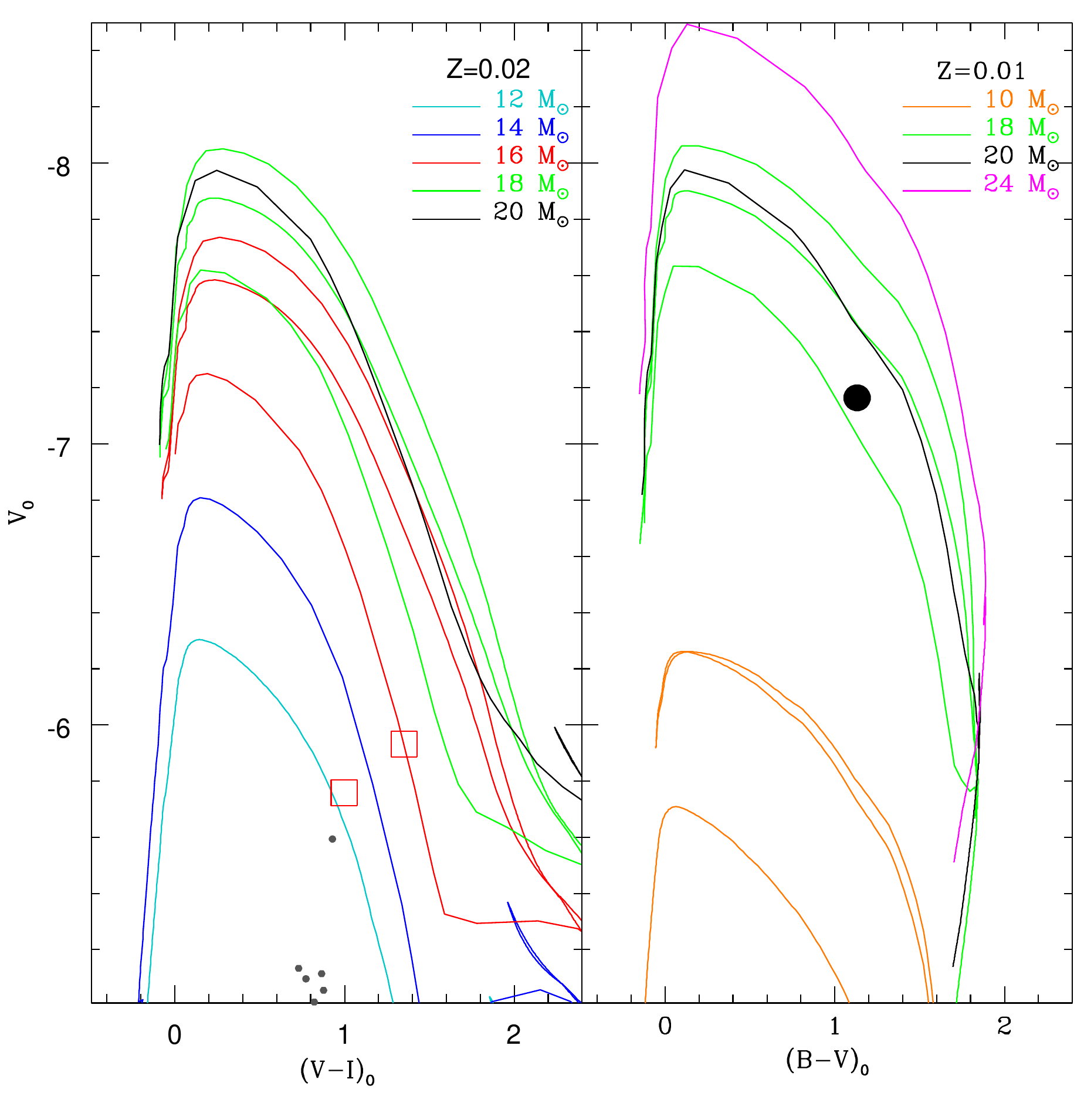}
    \caption{Left panel: $V_0$ versus $(V-I)_0$ CMD for the M31 ULPs (red open squares) by \citet{Ngeow+15}, 8-0326 and 8-1498, with stellar tracks for $Z=0.02$ and mass between 12 and 20 $M_{\odot}$ (see colors and labels in figure). Right panel: $V_0$ versus $(B-V)_0$ CMD for the M31 ULP (filled black circle) by \citet{Taneva+20} with stellar tracks for $Z=0.01$ and mass between 10 and 24 $M_{\odot}$ (see colors and labels in figure).
    }
    \label{fig:cmd_tracks_m31}
\end{figure}

In the left panel of Fig. \ref{fig:cmd_tracks_m31}, we plot  the $V_0$, $(V-I)_0$ CMD with the two M31 ULPs found by \citet{Ngeow+15}, 8-0326 and 8-1498 (red open squares), over-imposed to the evolutionary tracks for metallicity $Z=0.02$ and stellar masses ranging from 12 to 20 $M{\odot}$. In the right panel of the same figure, we plot the position of the M31 ULP by \citet{Taneva+20}, H42 (black filled circle), in  the $V$, $B-V$ CMD over-imposed to  the evolutionary tracks for  metallicity $Z=0.01$, and masses equal to 10, 18, 20 and 24 $M_{\odot}$.  
We notice that the ULP 8-0326 is located on the $M=16\ M_{\odot}$ evolutionary track, with $M$, $\log{L}$ and $T_e$ of  about 15.7$M_{\odot}$, 4.57 dex and 4300 K, respectively. From these values, we can infer the predicted pulsation period by relying on the period-luminosity-color-mass (PLCM) relation by \citet{DeSomma+20a} and find a period of about 91 d that is longer than the observed value \citep[74.427 d,][]{Ngeow+15}. To reproduce the observed period, we need to increase $T_e$ up to about 4500 K or decrease the luminosities down to about 4.43 dex. 
To preserve the mass-luminosity relation, also considering the non-negligible uncertainty related to color-temperature transformation \citep[with differences of about of 150 K due to variations in the adopted model atmospheres, see also][and references therein]{Marconi+15}, in the following we adopt  $T_e=$ 4500 K for ULP 8-0326 as starting point for our light curve model fitting procedure. 

As for  ULP 8-1498, its position in the CMD lies between the evolutionary tracks at 12 and 14 $M_{\odot}$. By interpolating between these mass values we expect that the mass, the luminosity level and the effective temperature are around 12.5 $M_{\odot}$, 4.3 dex and 5000 K, respectively. These values, when used as input parameters in the theoretical PLCM mentioned above,  correspond to a period of about 38 d, very different from the observed period of 83.181 d obtained by \citet{Ngeow+15}\footnote{We used \citet{Ngeow+15} data to re-determine the period of this ULP, obtaining the same result within the errors. On the contrary, adopting 38 d as period, the light curve is not phased.}. 

Concerning H42 ULP, \citet{Taneva+20} suggest a mass of 20 M$_{\odot}$. Indeed, from its position in the $V_0$, $(B-V)_0$ CMD, we infer a stellar mass ranging from 18 to 20 M$_{\odot}$. By assuming a mass in this range and the corresponding luminosity and effective temperature, as derived from the evolutionary tracks, we applied the PLCM relation by \citet{DeSomma+20a}. As a result, we derived a period around 100 days. To obtain a period of 177 days, we need to decrease the mass to about 10 M$_{\odot}$, but the corresponding magnitude and color are not consistent with the evolutionary track for this mass value (orange line in the right panel of Fig. \ref{fig:cmd_tracks_m31}).

Due to these inconsistencies between period mass and luminosity, we could not apply the light curve model fitting procedure to the variables 8-1498 and H42. Additional data are needed in order to understand if these results are affected by a poor period determination and to confirm the ULP nature of these variables.

To perform an accurate fit of 8-0326 light curves in $V$ and $I$ bands, we constructed a set of pulsation models with the  period equal to the observed one (within $\sim$2\%), a metal abundance  $Z=0.02$
(see Table \ref{tab:ULPsample}) and varying the physical parameters around the values obtained above from the comparison with the evolutionary tracks in the CMD. In particular, as a first step, we fixed the mass and built iso-periodic $Z=0.02$ model sequences varying the luminosity and in turn the effective temperature. Possible variations in the efficiency of super-adiabatic convection were also taken into account by varying the mixing length parameter $\alpha$ used to close the nonlinear system of equations in the hydrodynamical code \citep[see][for details]{Fiorentino+07,DeSomma+20a}.

The obtained theoretical light curves are transformed in the Johnson $V$ and $I$ bands, by adopting the atmospheric models by \citet{Castelli+97a,Castelli+97b}.

To constrain the quality of the fit on a quantitative basis, we adopted a $\chi^2$ analysis \citep[for details, see e.g.][]{Marconi+13,Marconi+17,Ragosta+19}. 

Once obtained the best combination of luminosity and effective temperature for which the predicted curves matches the observed ones, we fixed the effective temperature to the obtained value and built additional iso-periodic sequences varying the mass and, in turn, the luminosity. The best fit procedure was then repeated obtaining a final best fit model.
This procedure provides us with the intrinsic stellar parameters mass, luminosity, effective temperature together with the apparent distance modulus in all considered bands. 
Using the apparent distance moduli obtained from our procedure, and fitting the Cardelli law for absorption \citep{car89}, it is possible to derive the absolute distance modulus together with the absorption in the $V$ band.
To estimate the uncertainties on the fitted parameters, we performed a set of 1000 bootstrap simulations
consisting of resampling the photometric light curves and replicating the fit for every simulation (a detailed description of the fitting method is described in Molinaro et al., 2020, in prep.). 

\begin{figure}
	\includegraphics[width=\columnwidth]{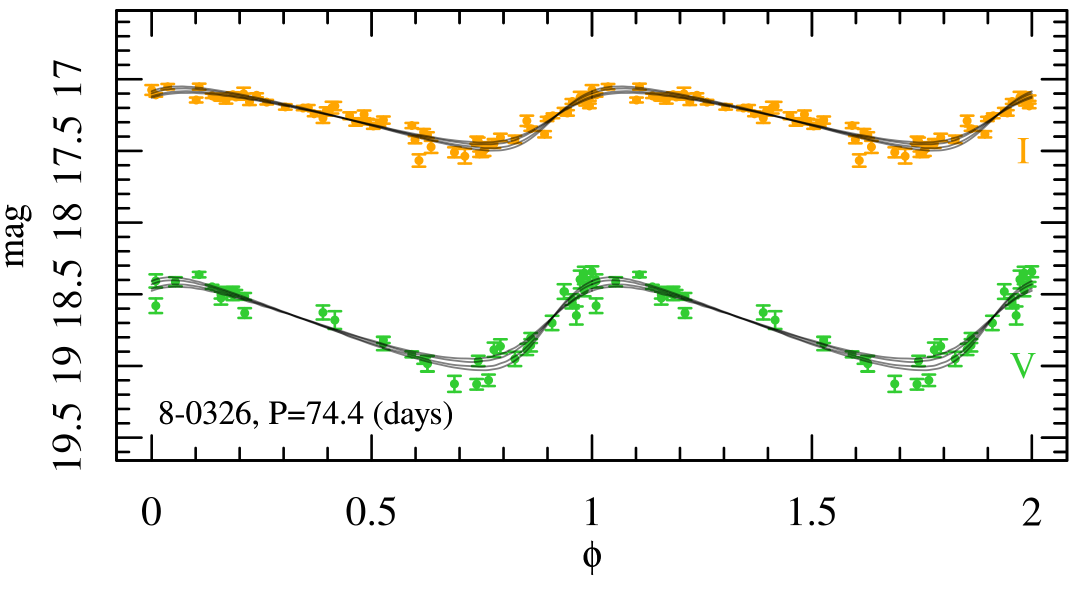}
    \caption{The ULPs light curves in $V$ and $I$ bands (green and yellow dots, respectively)   with the best fitting models (dark lines) in table \ref{tab:modelfit}.}
    \label{fig:ULPfit}
\end{figure}

In our analysis, for each pulsator,  we provide the best fit model obtained as a function of the adopted $\alpha$ parameter, and for each selection, additional three models that have a $\chi^2$ in agreement, within the errors, with the best fit model. These results are reported in Table \ref{tab:modelfit} that, for each  model, lists its $\alpha$ value, period $P_{mod}$, mass $M$, effective temperature $T_e$ and luminosity $\log(L/L_\odot)$, together with the associated $\chi^2$, the obtained distance modulus in the $V$ and $I$ bands, $\mu_V$ and $\mu_I$, the absorption in the $V$ band $A_V$ and the resulting absolute distance modulus $\mu_0$. Note that the uncertainties on the mass, effective temperature and luminosity are fixed equal to the parameter steps used to generate the grid of models. The observed ULP light curves with the corresponding best fitting model curves are shown in Fig. \ref{fig:ULPfit}. Our best estimates of the distance modulus and of the absorption are obtained as the weighted mean of the tabulated values: $\mu_0=24.124\pm0.012$ mag and $A_V=0.842\pm0.018$ mag. The inferred distance modulus is in agreement with the value obtained using the theoretical $VI$ Wesenheit for this ULP, but smaller than the M31 distance modulus by \citet{deGrijs+14} and \citet{PHAT15}. The absorption coefficient is in agreement within the errors with that obtained by \citet{Schlafly+11} $A_V=0.95\pm0.15$ mag. 

\begin{table*}
    \centering
    \caption{Direct fit results for LPV8-0326 (P. obs=74.4 days)}
    \label{tab:modelfit}
\begin{tabular}{cccccccccc}
\hline
$\alpha$ &  $P_{mod}$ & M & T & $\log(L/L_\odot)$ & $\chi^2$ & $\mu_V$ & $\mu_I$  & $A_V$ & $\mu_0$ \\ 
   $(dex)$ & $(d)$ & $(M_\odot)$ & $(K)$ & $(dex)$ & & $(mag)$ & $(mag)$ & $(mag)$ &$(mag)$ \\ \hline 
  1.59 &  76.46 & $15.9\pm0.2$ & $4630\pm25$ & $4.57\pm0.02$ & $0.841 \pm 0.113$  & $24.958 \pm 0.010$ & $24.640 \pm0.005$ & $0.828 \pm 0.027$ & $24.130 \pm 0.019$ \\ 
  1.60 &  76.46 & $15.9\pm0.2$ & $4630\pm25$ & $4.57\pm0.02$ & $0.856 \pm 0.127$ & $24.957 \pm 0.010$ & $24.641 \pm 0.005$ & $0.822 \pm 0.033$ & $24.135 \pm 0.024$  \\ 
  1.58 &  75.46 & $15.7\pm0.2$ & $4630\pm25$ & $4.57\pm0.02$ & $0.958 \pm 0.122$ & $24.954 \pm 0.009$ & $24.629 \pm 0.006$ & $0.847 \pm 0.026$ & $24.107 \pm 0.017$ \\ 
  1.56 &  76.08 & $15.7\pm0.2$ & $4680\pm 25$ & $4.58\pm0.02$ & $0.965 \pm 0.167$ & $25.006 \pm 0.011$ & $24.671 \pm 0.005$ & $0.871 \pm 0.032$ & $24.135 \pm 0.022$ \\ 
   \hline
\end{tabular}
\end{table*}

\section{Conclusions}

With the final aim of characterizing ULPs as standard candles able to reach in one step the Hubble flow, in this paper, we updated and enlarged the sample of 37 ULPs used by \citet{Fiorentino+12} and \citet{Fiorentino+13}. This new sample includes 62 objects{, for which we have $V$ and $I$ mean magnitudes,} and is composed by the 18 ULPs collected by \citet{Bird+09}, 2 ULPs in M81 \citep{Gerke+11}, 2 ULPs in M31 \citep{Ngeow+15} and 40 ULPs identified in 14 galaxies by \citet{Riess+16} and \citet{Hoffmann+16}, who enlarged, updated and re-calibrated the CC samples observed in the framework of the SH0ES project \citep{Riess+11}. To this sample we also add a M31 ULP found by \citet{Taneva+20} for which we have $B$ and $V$ mean magnitudes. The covered metallicity $12 + \log(O/H)$ ranges from $\sim 7.2$ to 9.2 dex. The properties of these pulsating stars in the P$W$ plane and in the CMD have been compared with different CC samples in the LMC \citep[OGLE,][]{Sosz+15} and NGC4258 \citep{Riess+16} and with the theoretical metal dependent Wesenheit function by \citet{Fiorentino+07}. The results do not confirm the almost flat slope suggested by \cite{Bird+09}, as including all the ULPs with $\log P \leq 2.15$ we find $W_{VI} = -2.15 \log P -4.89$ with $\sigma=0.38$. To investigate how this result can be influenced by the adoption of a sample collected by different sources, we also derived the Wesenheit relation by using only the SH0ES ULPs that represent a photometrically homogeneous sample obtaining a more robust result for the P$W$ relation, namely $W_{VI} = -2.89 \log P -3.42$ con $\sigma=0.36$. The difference between this last relation and that obtained by \citet{Sosz+15} for the LMC CCs, in the ULP period range is less than $\sim 0.05$ mag. 
On this basis, we perform a fit adding the ULPs to the LMC CC sample, obtaining $W_{VI} = -3.30 \log P -2.62$ with $\sigma=0.15$, both using all the ULPs with $\log P < 2.15$ that including only the SH0ES sample, in perfect agreement with that obtained using only the LMC OGLE CC by \citet{Sosz+15}. This result, together with the location of the ULPs in the CMD, seems to support the hypothesis that these variable are the extension at higher mass and luminosity of CCs, even if with a larger spread. This effect could be intrinsic, but also due to photometric limitations, crowding and blending. To get firm conclusions we need additional accurate data providing us with a photometrically homogeneous ULP sample, covering a larger range in period and metallicity. Moreover, we need to extend the CC pulsational models up to 20$M_{\odot}$ to get theoretical information on the  Instability Strip.

For the two M31 ULPs by \citet{Ngeow+15}, 8-0326 and 8-1498, and for the one by \citet{Taneva+20}, H42, we have time-series data published, so that we can analyse in detail their properties. Among these 3 ULPs, 8-1498 and H42 present inconsistency between their position in the CMD and the measured periods, thus suggesting that further observations are needed to confirm their nature as ULP. As for 8-0326, we apply the theoretical light-curve  model fitting method, by using new ad hoc pulsation models based  on a non linear, non local, time-dependent convective code \citep{Natale+08,MarconiLC+13,Ragosta+19}. Through a $\chi^2$ analysis of an extended set of models with the period fixed to the observed value and a wide range of input parameters, we were able to constrain the intrinsic stellar properties and constrain both the individual distance modulus and absorption.  The obtained distance modulus is found to agree with that obtained using the $VI$ Wesenheit for this ULP, but smaller than recent estimates of M31 distance by \citet{deGrijs+14} and \citet{PHAT15}, whereas the absorption coefficient is in agreement within the error with the value obtained by \citet{Schlafly+11}.

\section*{Acknowledgements}

We acknowledge Istituto Nazionale di
Fisica Nucleare (INFN), Naples section, specific initiative QGSKY. 
 We acknowledge partial financial support from ’Progetto Premiale’ MIUR MITIC (PI B. Garilli) and the INAF Main Stream SSH program, 1.05.01.86.28. This work has
made use of the VizieR database, operated at CDS, Strasbourg, France.
We thank the anonymous referee for some useful comments that improved the content of this paper.

\section*{DATA AVAILABILITY STATEMENTS}
The data underlying this article are available in the article.




\bibliographystyle{mnras}
\bibliography{main} 


\bsp	
\label{lastpage}
\end{document}